\documentstyle[12pt]{article}
\newlength{\extraspace}
\newlength{\extraspaces}
\newcommand{\be}{\begin{equation}
\addtolength{\abovedisplayskip}{\extraspaces}
\addtolength{\belowdisplayskip}{\extraspaces}
\addtolength{\abovedisplayshortskip}{\extraspace}
\addtolength{\belowdisplayshortskip}{\extraspace}}
\newcommand{\ee}{\end{equation}}

\newcommand{\ba}{\begin{eqnarray}
\addtolength{\abovedisplayskip}{\extraspaces}
\addtolength{\belowdisplayskip}{\extraspaces}
\addtolength{\abovedisplayshortskip}{\extraspace}
\addtolength{\belowdisplayshortskip}{\extraspace}}
\newcommand{\ea}{\end{eqnarray}}

\newcommand{\nonu}{\nonumber \\[.5mm]}

\begin{document}
\addtolength{\baselineskip}{.7mm}

\thispagestyle{empty}

\begin{flushright}
STUPP--93--135 \\ July 1993
\end{flushright}
\vspace{.6cm}

\begin{center}
{\large{\bf{Higher Order Calculations in
Renormalization }\\[3mm]
{Group Approach to Matrix Models}}}\\[1.5cm]
{\sc Yukihisa Itoh \footnote
{e-mail: itoh@th.phy.saitama-u.ac.jp}}\\[15mm]
{\it Physics Department, Saitama University \\[2mm]
Urawa, Saitama 338, Japan}\\[20mm]

{\bf Abstract}\\[1cm]

{\parbox{13cm}{\hspace{5mm}
We study higher order approximations in the renormalization
group approach to matrix models.
We use constraint equations on the free energy
resulting from a freedom of field redefinitions
and obtain the effective beta function for a single coupling
constant to the fifth order.
The fixed point and the string susceptibility
exponent are shown to approach the values obtained in
the exact solution as the order of approximations
becomes higher.
}}

\end{center}

\noindent

\vfill

\newpage


Matrix models [1] give a discrete
version of two-dimensional quantum gravity.
By studying the double scaling limit of the matrix models [2],
one can obtain the exact solutions of two-dimensional
quantum gravity coupled to minimal conformal field theories
with central charges $c \le 1$.
These results are consistent with the results of
a continuum version of two-dimensional
quantum gravity [3].
For the central charge $c>1$, we can construct matrix
models [4], but cannot obtain exact solutions.
Moreover, for those cases we do not know whether there is
a continuum theory or not.
Br\'ezin and Zinn-Justin [5] have proposed a renormalization
group approach to the matrix models in order to
understand $c > 1$ cases.
One can check the validity of this approach by applying
it to $c \leq 1$ theories, since
the exact solutions are known for these cases.
In ref. [5], they obtained reasonable results for the first
order approximation in the case of $c<1$.
The renormalization group approach has been further
studied in several papers [6-10].
\par
The purpose of this paper is to study higher order
approximations in the renormalization group approach to
the matrix models.
We consider a one-matrix model corresponding
to the case $c = 0$.
It was noticed in ref. [9] that one has to use
a freedom of field redefinitions to obtain higher
order results.
The exact beta functions of the vector model
were obtained in this way [9].
Although it is difficult to
obtain exact results for the matrix models using the
renormalization group approach,
we can compute beta functions
perturbatively in a coupling constant.
We derive constraint equations on the free energy
resulting from the freedom of field redefinitions.
Using these equations we
obtain the effective beta function for a single coupling
constant to the fifth order.
The fixed point of the beta function and the string susceptibility
exponent are shown to approach the values obtained in
the exact solution of the model as the order of approximations
becomes higher.
\par


We shall study the following one-matrix model:
\be
Z_N (g)=\int d^{N^2} \phi_N \exp \left[
- \sum_{k \geq 1}^{\infty} \sum_{l \geq k}^{\infty}
{g_{k,l} \over {kl}} tr \phi_N^k tr \phi_N^l
-N \sum _{m \geq 1}^{\infty} {g_{0,m} \over m} tr \phi_N^m
\right],
\label{2-1}
\ee
where $\phi_N$ is an $N \times N$ hermitian matrix.
The terms
$g_{0,m} tr \phi_N^m\ (m \geq 1)$ generate the usual
discretizations of random surfaces and the terms $g_{k,l}
tr \phi_N^k tr \phi_N^l ~(k \geq 1,l \geq k)$ generate
two random surfaces touching each other at one point [11,12].
This model was proposed in ref. [11] to include
effects of higher order intrinsic curvature terms.
Following ref. [5], by integrating out one row and
one column of $(N+1) \times (N+1)$ matrix $\phi_{N+1}$,
we will see that
the partition function $Z_N (g)$ fulfills the equation
\be
Z_{N+1} (g)=\Gamma_N (g) Z_N (g+\delta g),
\label{2-2}
\ee
where $\Gamma_N (g)$ is a factor which is a function
of $N$ and $g$.
{}From this relation we have a differential equation
\ba
{\partial \over \partial N} \ln Z_N (g)
&\!\!\! = &\!\!\!
\ln \Gamma_N (g)+ \sum_{k \geq 1}^{\infty}
\sum_{l \geq k}^{\infty} \delta g_{k,l}
{\partial \over \partial g_{k,l}} \ln Z_N (g)
\nonu
&\!\!\! &\!\!\!
+ \sum_{m \geq 1}^{\infty} \delta g_{0,m}
{\partial \over \partial g_{0,m}} \ln Z_N (g).
\label{2-3}
\ea
\par
Now, it is convenient to introduce rescaled
coupling constants [9],
\be
\tilde g_{k,l}={g_{k,l} \over g_{0,2}^{k+l \over 2}}
\qquad (k \geq 1, ~l \geq k),
\label{2-4}
\ee
and
\be
\tilde g_{0,m}={g_{0,m} \over g_{0,2}^{m \over 2}}
\qquad (m \geq 1).
\label{2-5}
\ee
By this rescaling, the quadratic term in the potential
becomes the standard form ${1 \over 2} tr \phi_N^2$, i.e.,
$\tilde g_{0,2} =1$.
It is clear that we have a relation
\be
Z_N (g) =(g_{0,2})^{-{N^2 \over 2}} Z_N (\tilde g),
\label{2-6}
\ee
where $Z_N (g)$ and $Z_N (\tilde g)$ are shorthand notations
of $Z_N (g_{k,l}\;(k \geq 1,~l \geq k),~ g_{0,m}\;(m \geq 1))$
and $Z_N (\tilde g_{k,l}\;(k \geq 1,~ l \geq k),~
\tilde g_{0,2}=1,~ \tilde g_{0,m}\;(m=1,3,\cdots))$
respectively.
We denote the partial derivatives with respect to $\tilde g_{k,l}
\;(k \geq 1, l \geq k)$, $\tilde g_{0,m}\;(m=1,3,\cdots)$
and $g_{0,2}$ by $\mid_{\tilde g}$, and those with
$ g_{k,l}\;(k \geq 1,~ l \geq k)$ and $g_{0,m}\;(m \geq 1)$
by $\mid_g$. They are related as
\be
\left. {\partial \over \partial g_{0,2}} \right|_g
=\left. {\partial \over \partial g_{0,2}} \right|_{\tilde g}
- \sum_{m=1,3,\cdots}^{\infty}
{m \over 2}
{\tilde g_{0,m} \over g_{0,2}} \left. {\partial \over \partial
\tilde g_{0,m}} \right|_{\tilde g}
- \sum _{k \geq 1}^{\infty} \sum_{l \geq k}^{\infty}
{k+l \over 2} {\tilde g_{k,l} \over g_{0,2}}
\left. {\partial \over \partial \tilde g_{k,l}} \right|_{\tilde g},
\label{2-7}
\ee
\be
\left. {\partial \over \partial g_{k,l}} \right|_g
={1 \over g_{0,2}^{k+l \over 2}}
\left. {\partial \over \partial \tilde g_{k,l}} \right|_{\tilde g}
\qquad (k \geq 1,~l \geq k),
\label{2-8}
\ee
\be
\left. {\partial \over \partial g_{0,m}} \right|_g
={1 \over g_{0,2}^{m \over 2}}
\left. {\partial \over \partial \tilde g_{0,m}} \right|_{\tilde g}
\qquad (m =1,3,\cdots).
\label{2-9}
\ee
If we define the free energy
\be
F(N,\tilde g)=-{1 \over N^2} \ln Z_N (\tilde g),
\label{2-10}
\ee
the renormalization group equation (\ref{2-3}) becomes
\ba
&\!\!\! &\!\!\!
\left[
N {\partial \over \partial N}
- \sum_{k \geq 1}^{\infty} \sum_{l \geq k}^{\infty}
\tilde \beta_{k,l} (\tilde g)
\left.
{\partial \over \partial \tilde g_{k,l}}
\right|_{\tilde g}
- \sum_{m=1,3,\cdots}^{\infty}
\tilde \beta_{0,m} (\tilde g)
\left.
{\partial \over \partial \tilde g_{0,m}}
\right|_{\tilde g}
+2
\right]
F(N,\tilde g)
\nonu
&\!\!\! &\!\!\!
=
-\ln g_{0,2}
-{1 \over N} \ln \Gamma_N
+ {\delta g_{0,2} N \over 2g_{0,2}},
\label{2-11}
\ea
where the $\tilde \beta$ functions are
\be
\tilde \beta_{k,l} (\tilde g)
=N
\left(
{\delta g_{k,l} \over g_{0,2}^{k+l \over 2}}
-{k+l \over 2}{\delta g_{0,2} \tilde g_{k,l} \over g_{0,2}}
\right)
\qquad (k \geq 1,~l \geq k),
\label{2-12}
\ee
\be
\tilde \beta_{0,m} (\tilde g)
=N
\left(
{\delta g_{0,m} \over g_{0,2}^{m \over 2}}
-{m \over 2}{\delta g_{0,2} \tilde g_{0,m} \over g_{0,2}}
\right)
\qquad (m=1,3,\cdots).
\label{2-13}
\ee


Now we shall consider the simplest case of
a cubic potential
\be
g_{0,2} \ne 0,
{}~~ g_{0,3} \ne 0,
{}~~ g_{0,m}=0 \quad(m=1,4,\cdots),
{}~~ g_{k,l}=0 \quad(k \geq 1, l \geq k).
\label{3-1}
\ee
The shifts $\delta g$ in eq. (\ref{2-2})
can be obtained exactly as follows.
The partition function $Z_{N+1} (g)$ is defined by
\be
Z_{N+1} (g)=\int d^{(N+1)^2} \phi_{N+1}
\exp
\left[
-(N+1)
\left(
{g_{0,2} \over 2} tr \phi_{N+1}^2
+{g_{0,3} \over 3} tr \phi_{N+1}^3
\right)
\right].
\label{3-2}
\ee
The matrix $\phi_{N+1}$ is parametrized in terms of a
submatrix $\phi_N$, a complex vector $v_N$ and a
real number $\alpha$:
\be
\phi_{N+1}
=\left(\matrix{\phi_N \hfill & v_N \hfill
\cr v_N^{*} \hfill & \alpha \hfill}\right).
\label{3-3}
\ee
Since all the terms involving $\alpha$
are of relative order $1 / N$ [5], we can set $\alpha =0$.
Hence we have
\ba
Z_{N+1} (g)
&\!\!\!=&\!\!\!
\int d^{N^2} \phi_N d^N v_N d^N v_N^{*}
\nonu
&\!\!\! &\!\!\!
\times \exp
\left[
-(N+1)
\left(
{g_{0,2} \over 2} tr \phi_N^2
+{g_{0,3} \over 3} tr \phi_N^3
+g_{0,2} v_N^{*} v_N
+g_{0,3} v_N^{*} \phi_N v_N
\right)
\right].~~~~
\label{3-4}
\ea
Since the $v_N$ and $v_N^{*}$ integrals are Gaussian,
we can evaluate them exactly.
After rescaling of $\phi_N$ we obtain
\ba
Z_{N+1} (g)
&\!\!\!=&\!\!\!
\{ -(N+1)g_{0,2} \} {}^N
\int d^{N^2} \phi_N
\exp
\left[
-N
\left(
{g_{0,2} \over 2} tr \phi_N^2
+{g_{0,3} \over 3} tr \phi_N^3
\right)
\right]
\nonu
&\!\!\! &\!\!\!
\times
\exp
\left[
- \sum_{k \geq 1}^{\infty} \sum_{l \geq k}^{\infty}
{\delta g_{k,l} \over {kl}} tr \phi_N^k tr \phi_N^l
-N \sum _{m \geq 1}^{\infty}
{\delta g_{0,m} \over m} tr \phi_N^m
\right],
\label{3-5}
\ea
where
\ba
&\!\!\! &\!\!\!
\delta g_{k,l}=0 \quad (k \geq 1,~l \geq k),
{}~~ \delta g_{0,2}={1 \over N}
\left( g_{0,2}-
\left( {g_{0,3} \over g_{0,2}}
\right)^2
\right),
\nonu
&\!\!\! &\!\!\!
\delta g_{0,3}={1 \over N}
\left( g_{0,3}+
\left( {g_{0,3} \over g_{0,2}}
\right)^3
\right),
{}~~ \delta g_{0,m} =-{1 \over N}
\left(
- {g_{0,3} \over g_{0,2}}
\right)^m \quad (m=1,4,\cdots).
\label{3-6}
\ea
The factor in eq. (\ref{2-2}) is given by
$\Gamma_N (g) = \{ -(N+1)g_{0,2} \} {}^N$.
We can also consider cases of other potentials,
but then the $v_N$ and $v_N^{*}$ integrals cannot be
evaluated exactly.
{}From eqs. (\ref{2-11}) and (\ref{3-6}), we have the following
renormalization group equation for the cubic potential
\be
\left[
N {\partial \over \partial N}
- \sum_{m=1,3,\cdots}^{\infty}
\tilde \beta_{0,m}
\left.
{\partial \over \partial \tilde g_{0,m}}
\right|_{\tilde g}
+2
\right]
F(N,\tilde g)
=
- \ln g_{0,2}
-{1 \over N} \ln \Gamma_N
+ {1 \over 2}-{1 \over 2}{\tilde g_{0,3}}^2,
\label{3-7}
\ee
where $\tilde \beta$ functions are
\be
\tilde \beta_{0,3} (\tilde g_{0,3})
={1 \over 2} \tilde g_{0,3}
- {5 \over 2} \tilde g_{0,3}^3,
\label{3-8}
\ee
\be
\tilde \beta_{0,m} (\tilde g_{0,3})
=(-\tilde g_{0,3})^m \qquad (m=1,4,\cdots).
\label{3-9}
\ee


We find that even if one starts from the cubic potential,
after integrating one low and one column
one obtains infinite number of other higher order
terms $tr \phi_N^m$.
To do higher order calculations,
we must take into account these higher order terms.
However, one may be able to put the potential back
to a cubic form by appropriate redefinitions of $\phi_N$.
Such a freedom of field redefinitions was used in ref. [9]
to obtain the exact beta functions in the vector model.
\par
The freedom of field redefinitions can be expressed as
constraint equations on the partition function.
These constraints were discussed in refs. [13, 14]
and were shown to have a Virasoro like structure.
In particular they were discussed before taking
a continuum limit in ref. [13].
They can be obtained by a change of integration variables
\be
\phi_N \to \phi_N + \delta \phi_N
\label{4-1}
\ee
in the partition function.
The transformation considered in ref. [13] are
\be
\delta \phi_N=\epsilon \phi_N^{n+1} \qquad (n \geq -1),
\label{4-2}
\ee
where $\epsilon$ is an infinitesimal parameter.
We call the constraints corresponding to the change of
variables (\ref{4-2}) as $L_n \ (n \geq -1)$.
In the case of our model (\ref{2-1}) the constraints
for the free energy $F(N, \tilde g)$ are

$L_{-1}$ condition:
\be
\left[
\left.
{\partial \over \partial \tilde g_{0,1}}
\right|_{\tilde g}
-3 {\tilde g_{0,3}}^2
\left.
{\partial \over \partial \tilde g_{0,3}}
\right|_{\tilde g}
\right]
F(N,\tilde g)
+\tilde g_{0,3}
=0,
\label{4-6}
\ee

$L_{0}$ condition:
\be
\left.
{\partial \over \partial g_{0,2}}
\right|_{\tilde g}
F(N,\tilde g)
=0,
\label{4-7}
\ee

$L_{1}$ condition:
\be
\left[
3
\left.
{\partial \over \partial \tilde g_{0,3}}
\right|_{\tilde g}
+4 \tilde g_{0,3}
\left.
{\partial \over \partial \tilde g_{0,4}}
\right|_{\tilde g}
-2
\left.
{\partial \over \partial \tilde g_{0,1}}
\right|_{\tilde g}
\right]
F(N,\tilde g)
=0,
\label{4-8}
\ee

$L_{2}$ condition:
\be
\left[
4
\left.
{\partial \over \partial \tilde g_{0,4}}
\right|_{\tilde g}
+5 \tilde g_{0,3}
\left.
{\partial \over \partial \tilde g_{0,5}}
\right|_{\tilde g}
-
\left.
{\partial \over \partial \tilde g_{1,1}}
\right|_{\tilde g}
+6 \tilde g_{0,3}
\left.
{\partial \over \partial \tilde g_{0,3}}
\right|_{\tilde g}
\right]
F(N,\tilde g)
-2
=0,
\label{4-9}
\ee

$L_{n} ~~(n \geq 3)$ conditions:
\ba
&\!\!\! &\!\!\!
\left[
(n+2)
\left.
{\partial \over \partial \tilde g_{0,n+2}}
\right|_{\tilde g}
+(n+3) \tilde g_{0,3}
\left.
{\partial \over \partial \tilde g_{0,n+3}}
\right|_{\tilde g}
\right.
\nonu
&\!\!\! &\!\!\!
\left.
-\sum_{a \geq 1,b \geq 1}^{a+b=n} ab
\left.
{\partial \over \partial \tilde g_{a,b}}
\right|_{\tilde g}
-2n
\left.
{\partial \over \partial \tilde g_{0,n}}
\right|_{\tilde g}
\right]
F(N,\tilde g)
=0.
\label{4-10}
\ea
Here we have set all the coupling constants to be zero
except $g_{0,2}$ and $g_{0,3}$ as in eq. (\ref{3-1})
after the differentiations.
\par
To do higher order calculations we will need
constraints derived from more general transformations of
$\phi_N$. Let us consider changes of variables
\be
\delta \phi_N={\epsilon \over N} \phi_N^{n+1}
tr \phi_N^p \qquad (n=-1,~0,~1,~p \geq 1).
\label{4-3}
\ee
The constraints $L_{n,p}\ (n=-1,~0,~1,~p \geq 1)$
corresponding to these changes of variables are
found to be

$L_{-1,p} ~(p \geq 1)$ conditions:
\be
\left[
\left.
{\partial \over \partial \tilde g_{1,p}}
\right|_{\tilde g}
+2 \tilde g_{0,3}
\left.
{\partial \over \partial \tilde g_{2,p}}
\right|_{\tilde g}
\right]
F(N,\tilde g)
=0,
\label{4-11}
\ee

$L_{0,p} ~(p=1,3,\cdots)$ conditions:
\be
\left[
2
\left.
{\partial \over \partial \tilde g_{2,p}}
\right|_{\tilde g}
+3 \tilde g_{0,3}
\left.
{\partial \over \partial \tilde g_{3,p}}
\right|_{\tilde g}
-
\left.
{\partial \over \partial \tilde g_{0,p}}
\right|_{\tilde g}
\right]
F(N,\tilde g)
=0,
\label{4-12}
\ee

$L_{0,2}$ condition:
\be
\left[
2
\left.
{\partial \over \partial \tilde g_{2,2}}
\right|_{\tilde g}
+3 \tilde g_{0,3}
\left.
{\partial \over \partial \tilde g_{2,3}}
\right|_{\tilde g}
+{3 \over 2} \tilde g_{0,3}
\left.
{\partial \over \partial \tilde g_{0,3}}
\right|_{\tilde g}
\right]
F(N,\tilde g)
- {1 \over 2}
=0,
\label{4-13}
\ee

$L_{1,p} ~(p \geq 1)$ conditions:
\be
\left[
3
\left.
{\partial \over \partial \tilde g_{3,p}}
\right|_{\tilde g}
+4 \tilde g_{0,3}
\left.
{\partial \over \partial \tilde g_{4,p}}
\right|_{\tilde g}
-2
\left.
{\partial \over \partial \tilde g_{1,p}}
\right|_{\tilde g}
\right]
F(N,\tilde g)
=0.
\label{4-14}
\ee
\par
We could consider more general transformations of the
form (\ref{4-3}) with $n \geq 2,~p \geq 1$ or
\be
\delta \phi_N=\epsilon \phi_N^{n+1}
{tr \phi_N^{p_1} \over N}{tr \phi_N^{p_2} \over N}
{tr \phi_N^{p_3} \over N} \cdots
\quad (n \geq -1,~p_1, p_2, p_3, \cdots \geq 1).
\label{4-5}
\ee
However, these transformations induce terms
$tr \phi^{k_1} tr \phi^{k_2} tr \phi^{k_3}
\cdots\ (1 \leq k_1 \leq k_2 \leq k_3 \leq \cdots)$
in the potential. So we have to begin with a more general
potential than (\ref{2-1}), which includes terms of the form
$tr \phi^{k_1} tr \phi^{k_2} tr \phi^{k_3} \cdots$.
To the order we will consider in this paper
we do not need to use such general transformations.
\par
Let us discuss implications of the above constraint equations.
The $L_{-1}$ condition (\ref{4-6}) shows that
$\left. {\partial \over \partial \tilde g_{0,1}} \right|_{\tilde g} F$
can be rewritten by
$\left. {\partial \over \partial \tilde g_{0,3}} \right|_{\tilde g} F$.
This means that the term $tr \phi_N$ in the potential
can be eliminated by an appropriate field redefinition.
The meaning of the $L_0$ condition (\ref{4-7}) is clear from the
definition of the free energy $F(N,\tilde g)$.
The $L_1$ condition (\ref{4-8}) shows that
$\left. {\partial \over \partial \tilde g_{0,4}} \right|_{\tilde g} F$
is also written by
$\left. {\partial \over \partial \tilde g_{0,3}} \right|_{\tilde g} F$.
The $L_2$ condition (\ref{4-9}), however,  shows
that we must introduce a new nonlinear term
$\left. {\partial \over \partial \tilde g_{1,1}} \right|_{\tilde g} F$
to rewrite
$\left. {\partial \over \partial \tilde g_{0,5}} \right|_{\tilde g} F$.
Similarly, from the conditions (\ref{4-10}),
$\left. {\partial \over \partial \tilde g_{0,n}} \right|_{\tilde g} F$
$(n \geq 6)$ are not given only by
$\left. {\partial \over \partial \tilde g_{0,3}} \right|_{\tilde g} F$
but also by nonlinear terms
$\left. {\partial \over \partial \tilde g_{a,b}} \right|_{\tilde g} F$
$(a+b = n-3)$.
By using $L_{n,p}$ conditions (\ref{4-11})-( \ref{4-14}),
these nonlinear terms can be rewritten by
$\left. {\partial \over \partial \tilde g_{0,3}} \right|_{\tilde g} F$
and terms higher order in $\tilde g_{0,3}$.
\par


We can now calculate the fixed point $\tilde g_{0,3}^{*}$
and the string susceptibility exponent $\gamma_1$
using the above results.
We assume that $\tilde g_{0,3}$ is small and use a
perturbation theory in $\tilde g_{0,3}$.
This assumption is justified if a value of the fixed point
obtained in this way is small.
\par
We begin with the first order calculation,
which taking into account up to and including terms of
order $\tilde g_{0,3}^3$ in the
renormalization group equation (\ref{3-7}).
Since $\tilde \beta_{0,m}$ $(m \geq 4)$
are of order $\tilde g_{0,3}^m$, we can
ignore them in the renormalization group
equation. Using the $L_{-1}$ condition, we can
rewrite $\tilde \beta_{0,1}$ term as the same form as
$\tilde \beta_{0,3}$ term.
Thus we obtain an effective renormalization group equation
with a single beta function $\tilde \beta_{0,3}^{eff(1)}$
\be
\left[
N {\partial \over \partial N}
-\tilde \beta _{0,3}^{eff(1)}
\left.
{\partial \over \partial \tilde g_{0,3}}
\right|_{\tilde g}
+2
\right]
F(N,\tilde g)
= -\ln g_{0,2}
- {1 \over N} \ln \Gamma_N
+ {1 \over 2}-{3 \over 2} \tilde g_{0,3}^3 ,
\label{5-1}
\ee
\be
\tilde \beta_{0,3}^{eff(1)} (\tilde g_{0,3})
=-{1 \over 2} \tilde g_{0,3}
+{11 \over 2} \tilde g_{0,3}^3.
\label{5-2}
\ee
The fixed point $\tilde g_{0,3}^*$ is determined by
$\tilde \beta_{0,3}^{eff} (\tilde g_{0,3}^*) =0$
and the string susceptibility exponent is given by
$\gamma_1 = {2 \over \tilde \beta{'eff}_{0,3} (\tilde g_{0,3}^*)}$
as in ref. [5].
{}From eq. (\ref{5-2}), we find the fixed point
\be
\tilde g_{0,3}^{*(1)}=\sqrt{1 \over 11}
=0.301511 \cdots,
\label{5-3}
\ee
and the string susceptibility exponent
\be
\gamma_1^{(1)}=2.
\label{5-4}
\ee
On the other hand, from the exact solution [2]
we know that the double scaling limit is achieved
near the critical value
\be
\tilde g_{0,3}^{*(exact)}=\sqrt{1 \over 12 \sqrt{3}}
=0.2193456\cdots,
\label{5-5}
\ee
with the string susceptibility exponent
\be
\gamma_{1}^{(exact)}=2.5~.
\label{5-6}
\ee
This first order calculation is essentially
equivalent to the first order calculation in ref. [5] for
the matrix model with quartic potential.
\par
Next, for the second order approximation we take into
account up to and including terms of
order $\tilde g_{0,3}^4$.
Ignoring $\tilde \beta_{0,m}$
($m \geq 5$) and using the $L_{-1}$ and $L_1$ conditions
we obtain the effective renormalization group equation
\ba
\left[
N {\partial \over \partial N}
-\tilde \beta _{0,3}^{eff(2)}
\left.
{\partial \over \partial \tilde g_{0,3}}
\right|_{\tilde g}
+2
\right]
F(N,\tilde g)
&\!\!\!=&\!\!\!
-\ln g_{0,2}
- {1 \over N} \ln \Gamma_N
\nonu
&\!\!\! &\!\!\!
+ {1 \over 2}-{3 \over 2} \tilde g_{0,3}^3
+ {1 \over 2} \tilde g_{0,3}^4 ,
\label{5-7}
\ea
\be
\tilde \beta_{0,3}^{eff(3)} (\tilde g_{0,3})
=-{1 \over 2} \tilde g_{0,3}
+{25 \over 4} \tilde g_{0,3}^3.
\label{5-8}
\ee
Here we have ignored a term of order $\tilde g_{0,3}^5$.
{}From eq.(\ref{5-8}), we find the fixed point
\be
\tilde g_{0,3}^{*(2)}
=0.282842\cdots,
\label{5-9}
\ee
and the exponent
\be
\gamma_1^{(2)}=2.
\label{5-10}
\ee
\par
At the third order approximation we take into account
up to and including terms of order $\tilde g_{0,3}^5$
in the renormalization group equation (\ref{3-7}).
By using the $L_{-1}$, $L_1$, $L_2$, $L_{-1,1}$ and
$L_{-1,2}$ conditions we have the effective
renormalization group equation
\ba
&\!\!\! &\!\!\!
\left[
N {\partial \over \partial N}
-\tilde \beta _{0,3}^{eff(3)}
\left.
{\partial \over \partial \tilde g_{0,3}}
\right|_{\tilde g}
{\partial \over \partial N}
-\tilde \beta _{2,2}^{eff(3)}
\left.
{\partial \over \partial \tilde g_{2,2}}
\right|_{\tilde g}
+2
\right]
F(N,\tilde g)
\nonu
&\!\!\!=&\!\!\!
-\ln g_{0,2}
- {1 \over N} \ln \Gamma_N
+{1 \over 2}-{3 \over 2} \tilde g_{0,3}^3
+{13 \over 10} \tilde g_{0,3}^4 ,
\label{5-11}
\ea
where
\be
\tilde \beta_{0,3}^{eff(3)} (\tilde g_{0,3})
=-{1 \over 2} \tilde g_{0,3} +{137 \over 20} \tilde g_{0,3}^3
-{39 \over 10} \tilde g_{0,3}^5,
\label{5-12}
\ee
\be
\tilde \beta_{2,2}^{eff(3)} (\tilde g_{0,3})
= {4 \over 5} \tilde g_{0,3}^6.
\label{5-13}
\ee
Although a nonlinear effective beta function
$\tilde \beta_{2,2}^{eff(3)}$ appears in
eq. (\ref{5-11}), we may neglect
it because it is of order $\tilde g_{0,3}^6$.
Then from eq. (\ref{5-12}) we find the fixed point
\be
\tilde g_{0,3}^{*(3)}=0.276238 \cdots,
\label{5-14}
\ee
and the exponent
\be
\gamma_1^{(3)}=2.09517 \cdots.
\label{5-15}
\ee
\par
Similarly we can continue to improve the approximation.
At the fourth order, using the $L_{-1}$,
$L_1$, $L_2$, $L_3$, $L_{-1,1}$, $L_{-1,2}$, $L_{0,2}$
and $L_{0,3}$ conditions and ignoring terms of order
$\tilde g_{0,3}^7$ we obtain
the effective renormalization group equation
\ba
\left[
N {\partial \over \partial N}
-\tilde \beta _{0,3}^{eff(4)}
\left.
{\partial \over \partial \tilde g_{0,3}}
\right|_{\tilde g}
+2
\right]
F(N,\tilde g)
&\!\!\!=&\!\!\!
-\ln g_{0,2}
- {1 \over N} \ln \Gamma_N
\nonu
&\!\!\! &\!\!\!
+ {1 \over 2}-{3 \over 2} \tilde g_{0,3}^2
+{59 \over 30} \tilde g_{0,3}^4
+{1 \over 2} \tilde g_{0,3}^6 ,
\label{5-16}
\ea
\be
\tilde \beta_{0,3}^{eff(4)} (\tilde g_{0,3})
=-{1 \over 2} \tilde g_{0,3} +{147 \over 20} \tilde g_{0,3}^3
-{69 \over 10} \tilde g_{0,3}^5,
\label{5-17}
\ee
and the results
\be
\tilde g_{0,3}^{*(4)}=0.270249 \cdots,
\label{5-18}
\ee
\be
\gamma_1^{(4)}=2.15892 \cdots.
\label{5-19}
\ee
At the fifth order, using the $L_{-1}$,
$L_1$, $L_2$, $L_3$, $L_4$, $L_{-1,1}$, $L_{-1,2}$,
$L_{-1,3}$, $L_{0,2}$ and $L_{0,3}$  conditions
and ignoring terms of order $\tilde g_{0,3}^8$, we
obtain the effective renormalization group equation
\ba
\left[
N {\partial \over \partial N}
-\tilde \beta _{0,3}^{eff(5)}
\left.
{\partial \over \partial \tilde g_{0,3}}
\right|_{\tilde g}
+2
\right]
F(N,\tilde g)
&\!\!\!=&\!\!\!
-\ln g_{0,2}
- {1 \over N} \ln \Gamma_N
\nonu
&\!\!\! &\!\!\!
+ {1 \over 2}-{3 \over 2} \tilde g_{0,3}^2
+ {443 \over 210} \tilde g_{0,3}^4
+ {1 \over 2} \tilde g_{0,3}^6 ,
\label{5-20}
\ea
\be
\tilde \beta_{0,3}^{eff(5)} (\tilde g_{0,3})
=-{1 \over 2} \tilde g_{0,3}
+{1089 \over 140} \tilde g_{0,3}^3
-{723 \over 70} \tilde g_{0,3}^5
-{36 \over 7} \tilde g_{0,3}^7,
\label{5-21}
\ee
and the results
\be
\tilde g_{0,3}^{*(5)}=0.266948 \cdots,
\label{5-22}
\ee
\be
\gamma_1^{(5)}=2.25313 \cdots.
\label{5-23}
\ee
\par
Thus, as the order of approximations becomes higher,
resulting values of the fixed point and the string
susceptibility exponent approach the exact values
(\ref{5-5}), (\ref{5-6}).
To carry out much higher order calculations,
we have to use constraints corresponding to more
general transformations
(\ref{4-3}) with $(n \geq 2,~p \geq 1)$ and (\ref{4-5})
starting from more general potential than (\ref{2-1}).
\par

\bigskip

\noindent
{\bf Acknowledgments} \par
The author would like to thank Y. Tanii for useful discussions
and careful reading of the manuscript.

\newpage
%

\end{document}